\documentclass[twoside,twocolumn,9pt]{article}
\usepackage{extsizes}
\usepackage[super,sort&compress,comma]{natbib} 
\usepackage[version=3]{mhchem}
\usepackage[left=1.5cm, right=1.5cm, top=1.785cm, bottom=2.0cm]{geometry}
\usepackage{balance}
\usepackage{mathptmx}
\usepackage{sectsty}
\usepackage{graphicx} 
\usepackage{lastpage}
\usepackage[format=plain,justification=justified,singlelinecheck=false,font={stretch=1.125,small,sf},labelfont=bf,labelsep=space]{caption}
\usepackage{float}
\usepackage{fancyhdr}
\usepackage{fnpos}
\usepackage[english]{babel}
\addto{\captionsenglish}{}
\usepackage{array}
\usepackage{droidsans}
\usepackage{charter}
\usepackage[T1]{fontenc}
\usepackage[usenames,dvipsnames]{xcolor}
\usepackage{setspace}
\usepackage[compact]{titlesec}
\usepackage{hyperref}
\usepackage{amssymb}

\usepackage{epstopdf}

\begin{document}

\twocolumn[
\begin{@twocolumnfalse}
\noindent\Large{\textbf{Configuration-specific insight to single-molecule conductance and noise data revealed by principal component projection method}}\\

\noindent\large{Zolt\'an Balogh,\textit{$^{a,b}$} Gr\'eta Mezei,\textit{$^{a,b}$} Nóra Tenk,\textit{$^{a}$} Andr\'as Magyarkuti,\textit{$^{a}$} and Andr\'as Halbritter$^{\ast}$\textit{$^{a,b}$}}\vspace{0.6cm}

\textit{$^{a}$~Department of Physics, Institute of Physics, Budapest University of Technology and Economics, Műegyetem rkp. 3., H-1111 Budapest, Hungary. E-mail: halbritter.andras@ttk.bme.hu}\\
\textit{$^{b}$~ELKH-BME Condensed Matter Research Group, Műegyetem rkp. 3., H-1111 Budapest, Hungary.}\\
\vspace{0.6cm}

\noindent\normalsize{
We explore the merits of neural network boosted, principal-component-projection-based, unsupervised data classification in single-molecule break junction measurements, demonstrating that this method identifies highly relevant trace classes according to the well-defined and well-visualized internal correlations of the dataset. To this end, we investigate single-molecule structures exhibiting double molecular configurations, exploring the role of the leading principal components in the identification of alternative junction evolution trajectories. We show how the proper principal component projections can be applied to separately analyze the high or low-conductance molecular configurations, which we exploit in $1/f$-type noise measurements on bipyridine molecules. This approach untangles the unclear noise evolution of the entire dataset, identifying the coupling of the aromatic ring to the electrodes through the $\pi$ orbitals in two distinct conductance regions, and its subsequent uncoupling as these configurations are stretched. 
}
\end{@twocolumnfalse} \vspace{0.6cm}]

\section{Introduction}
The field of single-molecule electronics (SME) has developed a broad range of experimental methods to gain insight to the properties of the ultimate smallest conductors, where a single molecule bridges two metallic electrodes.\cite{Cuevas2010,Aradhya2013,Xin2019,Gehring2019_review,Evers2020_review} The creation of single-molecule nanowires heavily relies on the so-called break junction methods, where the fine mechanical rupture of a metallic nanowire is utilized first to establish a single-atom metallic nanojunction, and afterwards to contact a single-molecule between the two apexes of a broken atomic junction.\cite{Muller,Cuevas2010,Aradhya2013,Gehring2019_review,Xu2003} The basic output of the break junction measurements is a statistical ensemble of conductance vs. electrode separation traces, from which the fingerprints of the single-molecule configurations can be visualized on conductance histograms. This analysis is frequently extended by more delicate experimental methods, like force,\cite{Aradhya2013,Frei2011,Nef2012,Mohos2014,Magyarkuti2018} noise,\cite{Balogh2021,Djukic2006,Adak2015a,Karimi2016,Lumbroso2018,Magyarkuti2018,Tang2019,Wu2020,Yuan2021_review,Kim2021_review,Pan2022,Shein2022} thermoelectric power,\cite{Aradhya2013,Widawsky2012,Garcia2016_review} thermal conductance,\cite{Cui2017,Cui2019,Mosso2019} quantum conductance fluctuation\cite{Smit2002,Csonka2004} or superconducting subgap spectroscopy measurements.\cite{Scheer1998,Makk2008} The combination of such methods was found to be extremely efficient in resolving the fine details of the structural and electronic properties of single-molecule nanowires.

In spite of this broad range of diagnostic tools, single-molecule electronics still suffers from the diversity of the data, i.e. the formation of a broad range of various single-molecule configurations during the repeated opening and closing of the junction. Conductance histograms are sufficient to resolve the \emph{conductance of the most probable molecular configurations}. However, if a deeper insight is required, or further properties are also analyzed in parallel with the conductance data, it is useful to filter out the traces indeed reflecting the molecular configuration under focus, otherwise, the mixture of alternative configurations in the dataset may obscure the result of the analysis. This kind of data filtering is a traditional ingredient of break junction data analysis mostly relying on well-defined, but custom-created filtering algorithms.\cite{Makk2012a,Balogh2014,Inkpen2015,Huang2017} The recent progress of machine learning (ML) based data analysis has brought a novel perspective to the field of SME as well.\cite{William2022_review,Lemmer2016,Wu2017,Lauritzen2018,Cabosart2019,Bamberger2020,Huang2020,Fu2020,Fu2021,Magyarkuti2021}  However, in this field the widespread supervised ML approaches are less favored, as pre-labeled example traces are usually not available, and the manual selection of training traces is also unreliable and unwanted. Instead, the \emph{unsupervised} identification of the relevant trace classes is desired, which is indeed possible by proper ML approaches, including reference-free clustering methods, or auto-encoder-based clustering.\cite{Lemmer2016,Wu2017,Cabosart2019,Bamberger2020,Huang2020,Vladyka2020}

Here we examine an alternative unsupervised feature recognition approach relying on the \emph{internal correlations} of the dataset. This method utilizes the
cross-correlation analysis of the conductance traces.\cite{Halbritter2010,Makk2012,Aradhya2013b,Balogh2014,Magyarkuti2016,Magyarkuti2021} The leading principal components (PCs), i.e. the most important 
eigenvectors of the 2D correlation matrix identify the most relevant correlations in
the dataset offering a unique tool for the efficient identification of the relevant trace classes.
The principal component projection (PCP) method was first introduced and successfully applied to identify artificially mixed traces of different molecules.\cite{Hamill2018} Later, the authors of this paper demonstrated the sorting of molecular and tunneling traces by PCP method.\cite{Magyarkuti2020} In the latter work, it was also shown that the classification accuracy is significantly improved if the PCPs are only used to automatically generate the most characteristic traces for the two trace classes, which are then used as the training dataset of a simple neural network (NN).\cite{Magyarkuti2020} 

Inspired by the initial success of the PCP approach, here we further exploit this method via the analysis of room-temperature Au--4,4'-bipyridine (BPY)--Au\cite{Xu2003,Quek2009,Kamenetska2010,Aradhya2012,Baghernejad14,Isshiki2018,Magyarkuti2020,Mezei2020,Magyarkuti2021} and Au--2,7-diaminofluorene (DAF)--Au\cite{Magyarkuti2018} molecular nanowires. Both systems exhibit two different molecular configurations, i.e. alternative junction evolution trajectories are conceivable. We demonstrate that all the leading principal components provide highly relevant classification criteria, i.e. the PCP method is indeed an efficient and robust data filtering tool to recognize the alternative trace classes, which are mixed in the entire dataset. This can be harvested in more complex single-molecule measurements, where a configuration-specific insight into the data is desired. The latter is demonstrated by our PCP-aided, configuration-specific $1/f$-type noise measurements on BPY single-molecule junctions. These measurements investigate the scaling of $1/f$-type noise amplitude with the junction conductance, which is a specific device fingerprint distinguishing through-bond and through-space molecular coupling.\cite{Adak2015a,Balogh2021,Magyarkuti2018,Wu2020,Pan2022} In the latter case, a weak bond is involved in the transport, where tiny distance fluctuations exponentially convert to a significant conductance noise. This yields conductance independent relative conductance noise, $\Delta G/G$ similar to a tunnel junction, where the exponential dependence of the conductance on the distance, $G\sim \exp(-\beta\cdot d)$ converts to $\Delta G/G \sim \mathrm{const.}$ for a constant $\Delta d$ distance fluctuation.\cite{Adak2015a,Balogh2021} As a sharp contrast, in case of through-bond coupling a strong chemical bond fixes the molecule-apex distance, and therefore the noise is not related to distance fluctuations in the
bond, rather to close-by atomic fluctuations of the metallic apex.  
This phenomenon results in a definite increase of the relative conductance noise with decreasing junction conductance yielding a scaling exponent close to $\Delta G/G \sim G^{-0.5}$.\cite{Adak2015a,Balogh2021} 
Our noise analysis on BPY molecules yields a complex, unclear $\Delta G/G$ vs. $G$ dependency. The PCP-based configuration-specific noise analysis, however, fully clarifies the noise characteristics of the two distinct molecular configurations, providing a refined insight to the bond evolution. 

\section{Results and Discussion}
\begin{figure}[t]
\centering{\includegraphics[width=0.48\textwidth]{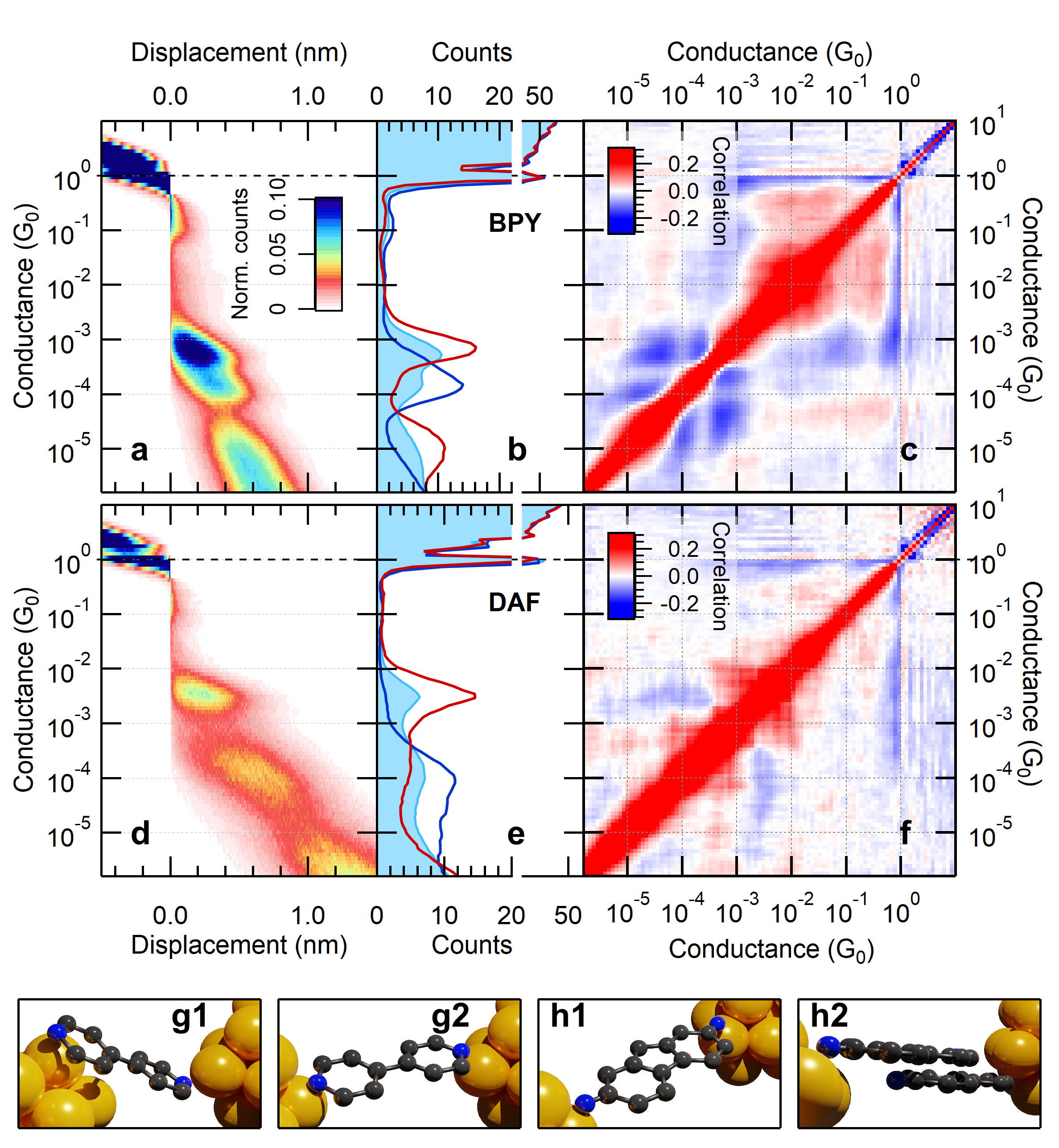}
\caption{{\bf Histograms and correlation plots of BPY and DAF molecules.} The top row represents the characteristics of BPY molecules, including the 2D histogram (a), the 1D histogram (blue area graph in panel (b)), and the  $C_{i,j}$ correlation plot (c). The red and blue lines in panel (b) represent the histograms of the data subsets obtained by EPCP analysis (see text). 
The second row similarly demonstrates the 2D histogram (d), the 1D histogram (e), and the correlation plot (f) for DAF molecules. On the 2D histograms the origin of the displacement axis is set 
to the position, where the conductance trace crosses $G_\mathrm{ref}=0.5\,$G$_0$. Panels g1 and g2 illustrate the hypothesized HighG and LowG BPY configurations, whereas h1 and h2 illustrate the HighG (monomer) DAF configuration and the LowG (dimer) DAF configuration.}
  \label{Fig1}}
\end{figure}

\textit{\textbf{Basics of the measurements.}} 
The break junction measurements are performed by scanning tunneling microscope (STM) or mechanically controllable break junction (MCBJ) methods, where the mechanically sharpened Au tip of a custom-designed STM setup is brought into contact with a gold thin film sample covered by the target molecules, or a gold wire is broken in molecular environment using a three-point bending geometry.\cite{Cuevas2010} In both cases, the rupture and the closing of the junction is repeated several thousand times, characterizing each rupture process by a conductance vs. electrode separation ($G(z)$) trace. From each measured trace, indexed with $r$, a single-trace histogram denoted by $N_i(r)=N(G_i,r)$ is
calculated by dividing the conductance axis to discrete bins ($G_i$) - and counting the number of data points in each bin. The conductance histogram of the entire dataset, $\left<N_i(r)\right>_r$ is obtained by averaging the single-trace histograms for all traces. These simple conductance histograms can be expanded to two-dimensional conductance histograms,\cite{Martin2008a,Quek2009} where the displacement information is also visualized. The such-obtained one-dimensional and two-dimensional conductance histograms for room-temperature Au-BPY-Au and Au-DAF-Au single-molecule junctions are presented in Figs. \ref{Fig1}a,b (BPY) and Figs. \ref{Fig1}d,e (DAF) (see the blue area graphs in the 1D histograms). In both cases, the formation of single-atom Au junctions is reflected by a sharp peak at the quantum conductance unit, $G_0=2e^2/h$, and both molecular systems exhibit double molecular configurations, one with higher and one with lower conductance (HighG/LowG). In the case of BPY, it is suggested that at a smaller electrode separation, the molecule can bind such that both the nitrogen linker and the aromatic ring is electronically coupled to the electrode, yielding a higher (HighG) single-molecule conductance (Fig.~\ref{Fig1}g1). Upon further pulling, the molecule slides to the apex and only the linkers couple to the electrodes (Fig.~\ref{Fig1}g2) yielding a
decreased (LowG) conductance value. \cite{Quek2009,Kamenetska2010,Aradhya2012,Kim2014b}
In the case of DAF the HighG/LowG configurations are interpreted as \emph{monomer}/\emph{dimer} molecular junctions (see Figs.~\ref{Fig1}h1,h2).\cite{Magyarkuti2018}

\textit{\textbf{Basics of principle-component-projection-based conductance trace classification.}} 
The one- and two-dimensional conductance histograms represent an average behavior of the conductance traces. However, further, highly relevant information is obtained from the correlation matrix of the dataset,\cite{Halbritter2010,Makk2012} highlighting the \emph{correlated deviations from the average behavior}:
$$C_{i,j}=C(G_i,G_j)=\frac{\left< (N_i(r)-\left< N_i(r)\right> )\cdot (N_j(r)-\left< N_j(r)\right> ) \right>}{\left<(N_i(r)-\left<N_i(r)\right>)^2\right>\cdot
\left<(N_j(r)-\left<N_j(r)\right>)^2\right>}.$$
The correlation matrices of the Au-BPY-Au and Au-DAF-Au systems are demonstrated in Figs.~\ref{Fig1}c,f, both exhibiting a rich correlation structure. Some aspects of the correlation matrix can be interpreted with proper experience,\cite{Halbritter2010,Makk2012,Aradhya2013b,Balogh2014,Magyarkuti2016,Magyarkuti2021} but the complexity of analyzing 2D correlation matrices is significantly reduced by concentrating on the \emph{most important correlations} represented by the $P_i^{(n)}$ principal components, i.e. the eigenvectors of the correlation matrix.\cite{Hamill2018} Here, the $n$ label sorts the principal components according to the descending order of the corresponding eigenvalue. 

The principal component projection of a certain conductance trace is obtained as $PCP^{(n)}(r)=\sum_i P_i^{(n)}\cdot N_i(r)$, i.e. the scalar product of the principal component and the single-trace histogram. According to the proposal of J. Hamill and coworkers, the $PCP^{(n)}(r) \lessgtr 0$ relation can be used to sort the traces to two relevant trace classes.\cite{Hamill2018} This idea was successfully applied to separate artificially mixed traces of two different molecular systems. It was also shown that the zero PCP value is not necessarily the proper threshold for the selection, and the accuracy of PCP-based data classification can be further refined by relying only on the traces with the largest positive/negative PCP (e.g. selecting $20\% - 20\%$ of all traces).\cite{Magyarkuti2020} If the classification of the \emph{entire} dataset is not targeted, solely the traces with these extreme principal component projection (EPCP) can be considered as a subset of the traces best representing some distinct behavior. Alternatively, these \emph{characteristic traces} can be applied as the training dataset for a simple double-layer feed-forward neural network.\cite{Magyarkuti2020} In this combined PCP-NN method the neural network \emph{learns} the important features from these training traces, and then generalizes for traces with less obvious characters. The decision-making process of this simplest possible neural network is well-represented by the so-called summed weight product,\cite{Magyarkuti2020} $SWP_i=\sum_j W^{(1)}_{i,j}\cdot W^{(2)}_{j}$, where $W^{(1)}_{i,j}$ 
/ $W^{(2)}_{j}$ is the neural weight matrix / vector between the input layer and the hidden layer / hidden layer and output neuron of the NN, respectively.
If $SWP_i$ is a large positive / negative number for a certain input, a large input value (i.e. a large histogram count) pushes the decision towards one or the other selection group. 

\textit{\textbf{Principle component projection analysis of room-temperature bipyridine and diaminofluorene single-molecule junctions.}} In our previous work the PCP-NN method was successfully applied in low-temperature single-molecule measurements,\cite{Magyarkuti2020} where the molecular pick-up rate was relatively low, and therefore, it was a primary task to separate molecular and non-molecular (tunneling) traces. In the following, we further demonstrate the merits and constraints of the EPCP and the combined PCP-NN method by analyzing the trace classes according to all the leading principal components in room-temperature measurements with BPY and DAF molecules. These measurements exhibit $\approx 100\%$ molecular pick-up rate, and therefore, instead of molecular/tunneling distinction, we rather focus on the analysis of the possible relations of the LowG and HighG molecular configurations. The results of the EPCP analysis are illustrated in Figs.~\ref{Fig1}b,e for a highly relevant principal component.
The red / blue curves show the conductance histograms for the traces with the largest positive / negative principal component projection, relying on $P^{(3)}$ ($P^{(2)}$) for BPY (DAF) molecules. All selections include $20\%$ of all traces, and for both molecules, one selection exhibits the dominance of the HighG configuration and the suppression of the LowG configuration, and the other selection behaves vice versa. The possibility of this separation is already surprising, as the 2D histograms of both molecular systems (Figs.~\ref{Fig1}a,d) suggest that first the HighG configuration appears, and the LowG configuration only forms upon further stretching. This scenario would imply that the LowG configuration appears together with the HighG one. As a very sharp contrast, the EPCP method demonstrates, that one can find a significant portion (in this case $20\%$) of the traces, where the pronounced LowG peak can be observed without any significant HighG weight. This means that the \emph{average behavior} represented by the 1D and 2D histograms may obscure possible highly relevant junction evolution trajectories, which can be, however, uncovered by the proper principal component projection.

\begin{figure}[!t]
\centering{\includegraphics[width=0.48\textwidth]{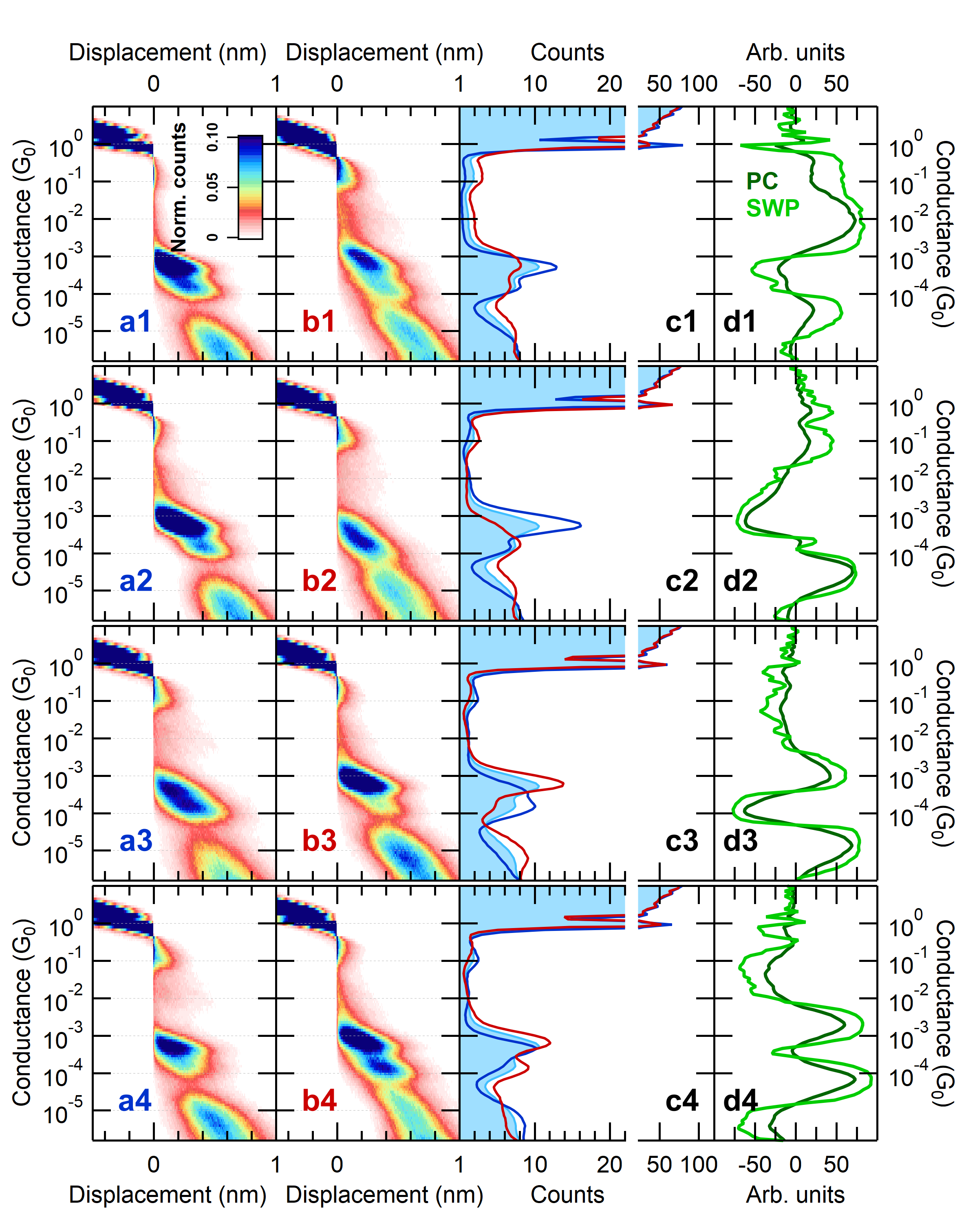}
\caption{{\bf PCP-NN analysis of Au-BPY-Au single-molecule junctions according to the four leading principal components.} The analysis according to $P^{(n)}$ is presented in the $n$th row, the corresponding principal components are shown by dark green lines in panels (d1-d4). The PCP-NN method classifies all the traces to two groups, trace class (a) and (b). The corresponding 2D histograms are respectively shown in panels (a1-a4) and (b1-b4), whereas the 1D histograms are shown in panels (c1-c4) by blue lines (trace class (a)) and by red lines (trace class (b)). As a reference, the blue area graphs represent the 1D histogram for the whole dataset. In panels (d1-d4) the light green lines represent the corresponding summed-weight-product of the neural network. The entire dataset includes 18301 traces, from which all the traces are classified into one or the other trace class. For the various principal components, the number of traces in a trace class varies in the range of $48.5\%-51.5\%$ of all traces.}
  \label{Fig2}}
\end{figure}

Further on, we extend the EPCP method by the neural network supplement and analyze the such obtained trace classes for the four leading principal components in room-temperature Au-BPY-Au junctions (Fig.~\ref{Fig2}). The rows of the figure respectively represent the analysis according to the first, second, third, and fourth principal components. For a certain principal component, the separated two trace classes are demonstrated both by 2D conductance histograms (Figs. \ref{Fig2}a1-a4 and b1-b4), and by 1D conductance histograms (Figs. \ref{Fig2}c1-c4). In the latter case, the light blue background histogram represents the reference histogram for the whole dataset, whereas the red and blue lines demonstrate the histograms of the two datasets obtained by the PCP-NN analysis. In the fourth column (Figs. \ref{Fig2}d1-d4) the actual principal component (dark green lines), and the summed-weight products of the neural network (light green lines) are demonstrated. Naturally, the selected one / other trace class exhibits pronounced weights in those conductance regions, where the PC and the SWP exhibit large positive / negative values. 

First of all, it is clear from Figs. \ref{Fig2}d1-d4 that the SWPs very much resemble the principal components for all the PCs. In the case of panels d2 and d3, the shape of the PC and the SWP is almost identical, whereas, in case of panels d1 and d4, the SWP refines and enhances the features of the PC, i.e. significantly sharper transitions are observed between the negative and positive regions. This already underpins the added value of the neural network extension of the simple PCP analysis: whereas the NN analysis basically learns the internal correlations of the dataset from the PCP, it is able to define more precise and sharper boundaries between the relevant conductance regions. Furthermore, we emphasize again that in the case of the simple PCP method, the best threshold PCP value between the two trace classes is ill-defined (it is not generally zero), but the NN extension is able to find the optimal boundary between the two trace classes. 

The analysis according to the first principal component enhances/suppresses both histogram peaks in the molecular region compared to the entire dataset. The background counts outside of the region of the molecular peaks behave vice versa, these are strongly suppressed/enhanced for the trace class (a)/(b) (see the blue and red lines in Fig. \ref{Fig2}c1, and the corresponding 2D histograms in Fig.~\ref{Fig2}a1/a2, where the blue/red label color matches the linecolor of the subset histograms in panel c1). Interestingly, the weight in the single-atom ($1\,$G$_0$) region and a bit higher conductance ($\approx 1.2-1.5\,$G$_0$) is also a significant part of the data filtering. For trace class (a)/(b) the $1\,$G$_0$ peak is significantly  enhanced/suppressed, whereas the weight at somewhat higher conductance is suppressed/enhanced, respectively. The latter phenomenon resembles the precursor effect reported in Ag-CO-Ag single-molecule junctions,\cite{Balogh2014} where the presence of the parallel-bound CO molecule spoils the formation of the clear  $1\,$G$_0$ single-atom configuration, and rather a broader weight is observed at somewhat larger conductance. Here, we argue, that well-defined BPY single-molecule junctions are best formed after the rupture of a well-defined single-atom Au junction with $1\,$G$_0$ conductance. Any contamination, that spoils the formation of this clean single-atom structure also hinders the formation of a well-defined Au-BPY-Au single-molecule structure, but rather enhances the background weight above the single-molecule and single-atom conductance. Therefore, the PCP-NN analysis according to the first principal component is efficient in separating \emph{ideal} molecular traces with well-defined Au-BPY-Au single-molecule configurations from \emph{obscured} molecular traces. The PCP-NN analysis  also highlights precursors of these trace classes in the $1\,$G$_0$ conductance region.

While $P^{(1)}$ filters the traces by suppressing or highlighting both molecular peaks, the further principal components perform a peak-specific classification. In case of $P^{(2)}$, the HighG molecular peak is either enhanced or suppressed and the LowG molecular peak is left unchanged (Fig.~\ref{Fig2}a2-c2); in case of $P^{(3)}$, either the HighG peak is enhanced and the LowG peak is suppressed, or vica versa (Fig.~\ref{Fig2}a3-c3); in case of $P^{(4)}$, the LowG peak is enhanced or suppressed, whereas the HighG peak is left more-or-less unchanged (Fig.~\ref{Fig2}a4-c4). Although the PCP-NN analysis does not uncover the physical origin of these various trace classes, it makes it very clear, that the \emph{average} and well-known junction evolution seen on the 2D conductance histogram of the BPY molecules (Fig. \ref{Fig1}a) is definitely not a single \emph{representative} junction evolution. While the former (Fig.~\ref{Fig1}a) implies the sequential formation of the HighG and LowG configurations on the \emph{same trace}, the PCP-NN-based data sorting very clearly highlights the presence of various fundamentally different junction evolution trajectories, which all represent a significant portion of the traces. More importantly, these possible significant junction evolution trajectories are not sorted according to some subjective manual data filtering criterion, nor by a black-box-like high-complexity machine learning method, rather they rely on the well-defined internal correlations of the dataset.   

\begin{figure}
  \centering{\includegraphics[width=0.48\textwidth]{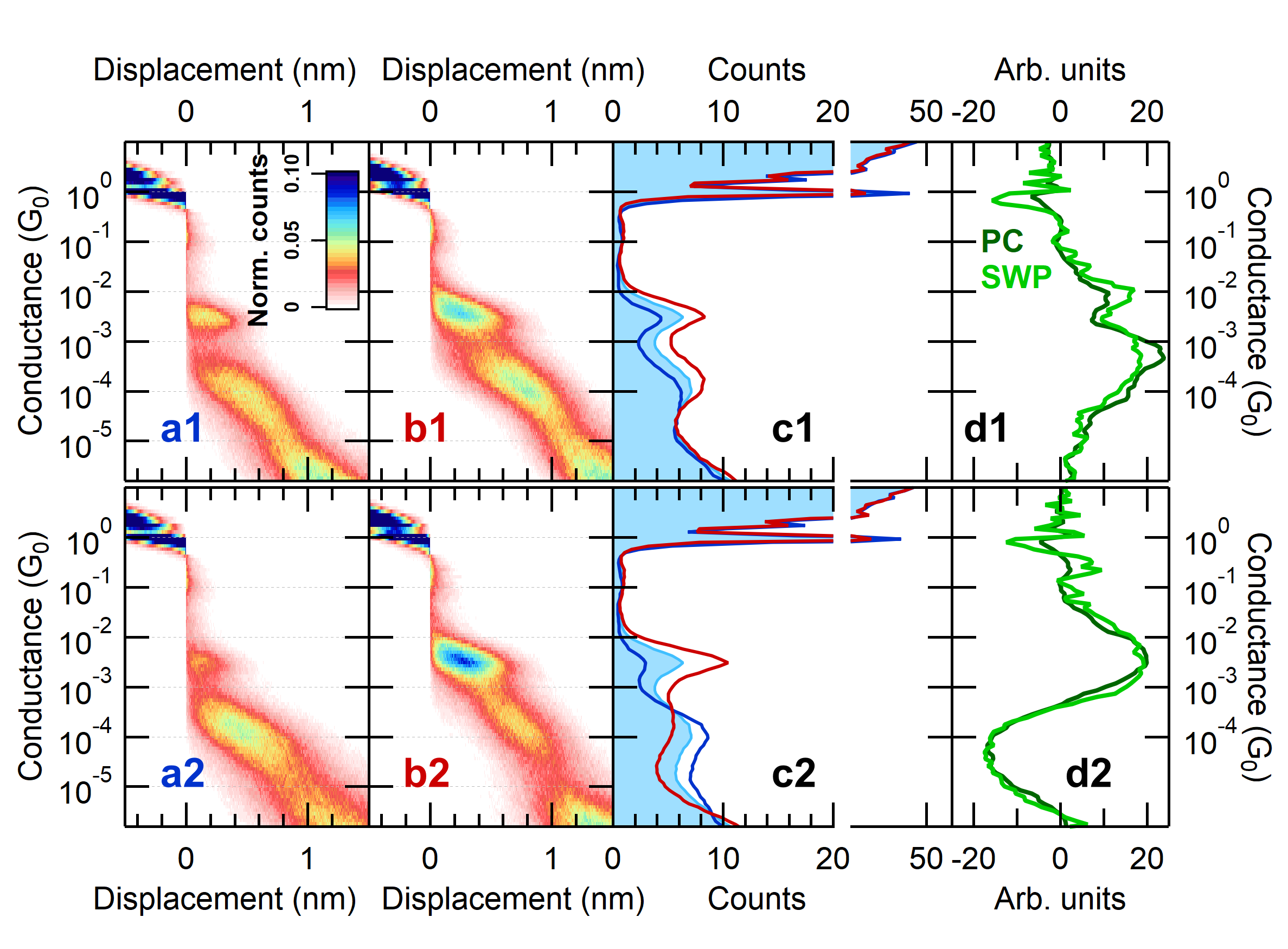}
  \caption{{\bf PCP-NN analysis of Au-DAF-Au single-molecule junctions.} The analysis according to $P^{(1)}$ and $P^{(2)}$ is respectively presented in the first and second row, the corresponding principal components are shown by dark green lines in panels (d1,d2). Panels (a1,a2) and (b1,b2) respectively demonstrate the 2D histograms of trace classes (a) and (b). The corresponding 1D histograms are respectively shown by blue and red lines in panels (c1,c2) in comparison to the histogram of the entire dataset (light blue area graph). In panels (d1,d2) the light green lines represent the corresponding summed-weight-product of the neural network. The entire dataset includes 7001 traces. The selected trace classes include $45\%-55\%$ of all traces.
}
  \label{Fig3}}
\end{figure}

Next, we perform a similar analysis on Au-DAF-Au single-molecule junctions (Fig.~\ref{Fig3}). The analysis according to $P^{(1)}$ (Fig.~\ref{Fig3}a1-d1) provides a somewhat similar result to the BPY molecules: trace class (a) yields the suppression of both molecular peaks, whereas trace class (b) enhances both molecular peaks. 
The effect of the projection to $P^{(2)}$ (Fig.~\ref{Fig3}a2-d2) is similar to the $P^{(3)}$ projection in case of the BPY molecules, i.e. in trace class (a) the HighG peak is suppressed, and the LowG peak is enhanced, and vice versa in trace class (b). In case of DAF molecules, the further principal component projections ($P^{(3)}$ and $P^{(4)}$) show very similar results to the $P^{(2)}$ projection. 

To demonstrate the robustness of the PCP-NN analysis, we have also investigated its sensitivity on the preparation of the data. First, we have found that independent measurements on Au-BPY-Au junctions using two completely different measurement methods (MCBJ/STM break junctions) yield very similar results, underpinning that the correlation plot and its principal components indeed act as specific fingerprints of the possible single-molecule junction evolution trajectories. On the other hand, the PCP-NN method is found to be especially sensitive to the temporal homogeneity of the data, inhomogeneous data portions may introduce severe artifacts to the analysis. These aspects are discussed in more detail in the ESI. 

\begin{figure}[t!]
\centering{\includegraphics[width=0.35\textwidth]{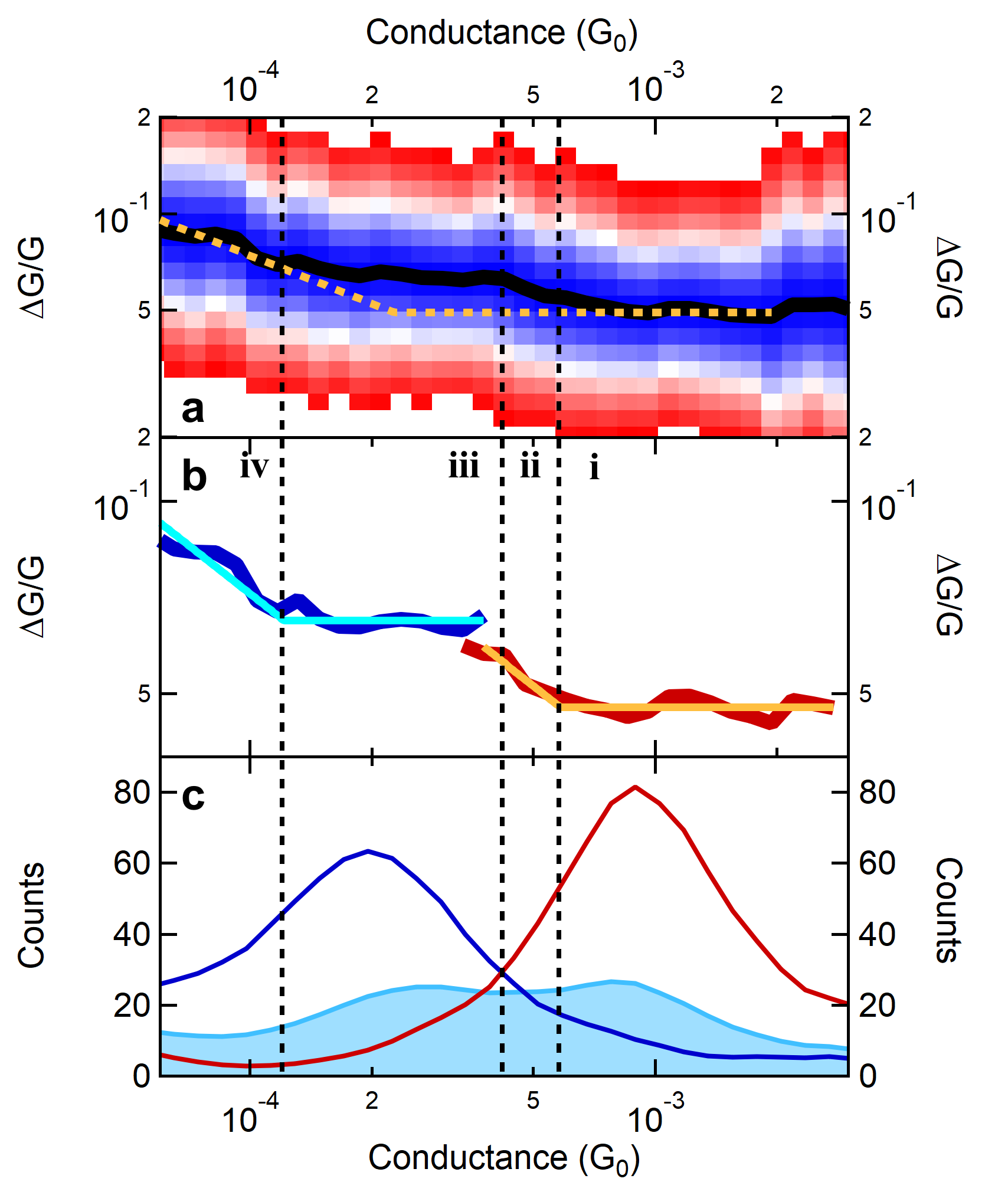}
  \caption{{\bf Configuration-specific $1/f$-type noise analysis of Au-BPY-Au junctions.} 
  (a) Most probable relative conductance noise as a function of the conductance (black line) for the entire dataset. The colorscale illustrates the scattering of the noise data (see ESI for more details). The orange dashed line represents the hypothetical noise evolution discussed in the text.
(b) Configuration-specific noise analysis demonstrating the noise evolution separately for the HighG (red) and LowG (blue) configurations. The orange and light blue lines represent the best-fitting model curves. (c) Conductance histogram of the entire dataset (light blue area graph) together with the configuration-specific EPCP projections (blue and red lines). The roman numbers together with the black dashed boundary lines illustrate the four characteristic regions of the junction evolution, as discussed in the text.}
  \label{Fig4}}
\end{figure}

\textit{\textbf{EPCP-aided interpretation of single-molecule $1/f$-type noise measurements.}} As mentioned in the introduction, the scaling of the $1/f$-type noise amplitude with the junction conductance supplies fundamental information about the nature of the metal-molecule bonds.\cite{Adak2015a,Magyarkuti2018,Wu2020,Balogh2021,Pan2022} Here, we investigate the noise characteristics of the widely-studied Au-BPY-Au single-molecule junctions. As an initial hypothesis, in the HighG configuration the aromatic ring also couples to the metallic apex which may introduce a significant through-space contribution to the transport, whereas the LowG configuration relies on the transport through the linkers, anticipating through-bond characteristics. The measured noise characteristics, i.e. the most probable relative conductance noise, $\Delta G/G$  is shown by black line in Fig.~\ref{Fig4}a for the entire dataset (see the light blue area graph in Fig.~\ref{Fig4}c for the corresponding conductance histogram in the molecular region, and the ESI for more details on the noise analysis). This curve indeed exhibits through-space-type $\Delta G/G \sim \mathrm{const.}$ characteristics at the top tail of the HighG peak, 
and through-bond-type $\Delta G/G \sim G^{-0.5}$  characteristics at the bottom tail of the LowG peak.
As a comparison, the orange dashed line represents the envisioned noise evolution according to the above discussed considerations, i.e. through-space-type noise in the entire HighG region and through-bond-type noise in the entire LowG region. More precisely, the orange dashed line extrapolates the best fitting $\Delta G/G \sim G^{-0.5}$ / $\Delta G/G \sim \mathrm{const.}$  tendencies at the low/high conductance \emph{sides} of the investigated conductance region to the entire conductance range assuming a sharp transition at the intersection of these lines. It is clear that in the intermediate conductance region 
the noise data clearly deviate from this supposed behavior.
This intermediate conductance region, however, is positioned around the conductance boundary between the HighG and LowG configurations ($G\approx 4\cdot 10^{-3}\,$G$_0$), where the largest diversity of the possible junction evolution paths appears according to the above PCP analysis.  Therefore, the interpretation of the unexpected noise evolution would remain highly speculative as long as the noise contributions of the two molecular configurations are not separated very clearly. 

The EPCP-based selection demonstrated by the red and blue histograms in Fig.~\ref{Fig1}b offers a unique possibility to very clearly separate the HighG and LowG configurations, investigating their noise characteristics \emph{separately}. The most probable $\Delta G/G$ values can be generated separately for these two data subsets, as demonstrated by the red and blue curves in Fig.~\ref{Fig4}b. According to our hypothesis, we would expect $\Delta G/G \sim \mathrm{const.}$ for the HighG subset (red) and $\Delta G/G \sim G^{-0.5}$ for the LowG subset (blue). This is, however, clearly contradicted by the data. Instead of the markedly \emph{different} noise characteristics of the HighG and the LowG configurations, both configurations exhibit very \emph{similar} noise evolution with through-space-type constant $\Delta G/G$ at the top part of the actual subset histogram peak, and through-bond-type characteristics ($\Delta G/G \sim G^{-0.5}$) at the bottom tail of the subset histogram peaks. The latter tendencies are also illustrated in Fig.~\ref{Fig4}b by the orange (light blue) line representing the best fitting model curves assuming \emph{constant}/$G^{-0.5}$ noise evolution above/below a certain conductance threshold  for HighG (LowG) configurations. Again, the deviation from the expected behavior is the most striking in the intermediate conductance region, where both the HighG and the LowG data subsets exhibit \emph{opposite} behavior to the initial hypothesis, i.e. through-bond-type behavior for the HighG configuration and through-space-type behavior for the LowG configuration. 

According to this configuration-specific analysis, we can identify four distinct conductance intervals according to distinct noise evolution characteristics (see the roman numbers in panel Fig.~\ref{Fig4}b and the separating dashed lines). (i) At the top part of the HighG peak the HighG configuration indeed exhibits additional $\pi$-coupling between the pyridine ring and the
electrodes, giving rise to constant through-space-like noise characteristics. (ii) As the HighG configuration is stretched (i.e. at the bottom tail of the HighG peak) this coupling vanishes giving rise to increasing, through-bond-type noise characteristics. (iii) At the top part of the LowG peak again a significant constant, through-space noise contribution is observed, indicating that the pyridine ring couples to the side of the electrode in this region as well. (iv) Finally, as the LowG configuration is stretched, this $\pi$-coupling vanishes again, the molecule straightens out, and again increasing, through-bond-type noise characteristics are observed (bottom tail of the LowG peak). More precisely, the borders between the i/ii and iii/iv intervals are identified by the points, where the fitted noise characteristics (orange and light blue lines in panel (b)) turn from constant to $\Delta G/G \sim G^{-0.5}$ dependency ($G=5.8\cdot 10^{-4}\,$G$_0$ for i/ii and $G=1.2\cdot 10^{-4}\,$G$_0$ for ii/iv), whereas the border between the ii/iii intervals is the crossing point of the HighG and LowG histograms in panel (c), $G=4.2\cdot 10^{-4}\,$G$_0$. Note, howevever, that these conductance regions rely on a rather simplifyed model, considering exactly $\sim G^{-0.5}$ and $\sim G^{0}$ scaling exponents with a sharp transition between these. It would be more realistic to permit deviations from these limiting scaling factors, and a more gradual transition between the two distinct behaviors, but due to the limitied resolution of the noise data we rather apply the most simple model grabbing the most important rough tendencies.

The above, peak-specific insight to the noise characteristics explores the cases, where solely the HighG or solely the LowG configurations appear. However, the noise evolution of the entire dataset (black curve in Fig.~\ref{Fig4}a) , where the two molecular configurations often appear sequentially on the same trace, follows the same trends as the separated configurations (through-space/through-bond/through-space/through-bond characteristics in regions i/ii/iii/iv, respectively). This suggests, that the above-described physical mechanisms are not specific to the separate, or the sequential appearance of the two configurations, rather a specific conductance interval (i-iv)  always shows similar noise evolution. In case of the sequential appearance of the two configurations, we argue that around the ii/iii boundary, the  molecule may jump to another bonding site, such that the pyridine ring again flips to the side of the electrode giving rise to a significant constant, through-space noise contribution. 

\section{Conclusions}
In conclusion, we have demonstrated the merits of PCP-based unsupervised data classification in single-molecule break junction measurements. According to our analysis on room-temperature Au-BPY-Au and Au-DAF-Au single-molecule structures, all the leading principal components project the conductance traces to highly relevant data subsets, signaling alternative junction evolution trajectories which are obscured on the 1D and 2D conductance histograms. 
These results demonstrate an efficient classification according to the well-established and well-visualized internal  correlations of the dataset, i.e. the PCP-method lacks the disadvantage of the black-box-like nature of more complex machine learning approaches, as well as the subjective nature of custom-created manual data filtering algorithms. 

Finally, we demonstrated that these efficient data classification capabilities can be exploited in the measurement of further physical quantities, where the mixture of alternative junction evolution trajectories would obscure the analysis. In particular, the PCP-method enabled us to perform a configuration-specific 1/f-type noise analysis, uncovering refined details of the junction evolution:
we identified the coupling of the aromatic ring to the electrodes through the $\pi$ orbitals in
two distinct conductance regions, and its subsequent uncoupling as these configurations are stretched.

\section*{Acknowledgements}
The authors acknowledge useful discussion with Latha Venkataraman on the basics of noise analysis, and with Gemma C. Solomon, Joseph M. Hamill, William Bro-Jørgensen and Kasper P. Lauritzen on machine-learning-based data analysis methods.  
This research was funded by the Ministry of Culture and Innovation and the National Research, Development and Innovation Office under Grant Nr. TKP2021-NVA-02 and the NKFI K143169 grant. Z.B. acknowledges the support of the Bolyai János Research Scholarship of the Hungarian Academy of Sciences and the \'UNKP-22-5 New National Excellence
Program of the Ministry for Innovation and Technology
from the source of the National Research, Development, and Innovation
Fund.

\section*{Author contributions}
The measurements were performed by G. Mezei and Z. Balogh (combined noise and conductance measurements) and by A. Magyarkuti and N. Tenk (traditional conductance measurements). The PCP analysis was performed by Z. Balogh, N. Tenk and G. Mezei, the neural network supplement was implemented by N. Tenk and A. Magyarkuti. The noise analysis was performed by G. Mezei and Z. Balogh. The idea of the project was conceived and the project was supervised by A. Halbritter, with a fundamental contribution of Z. Balogh in the noise measurement part. The manuscript was written by A. Halbritter and Z. Balogh with remarks from all authors. 

\balance

\bibliography{PCP_refs}
\bibliographystyle{arXiv}

\end{document}

\maketitle

\section{Sensitivity of the PCP analysis on the data preparation}

\begin{figure}[t!]
\centering{\includegraphics[width=0.7\textwidth]{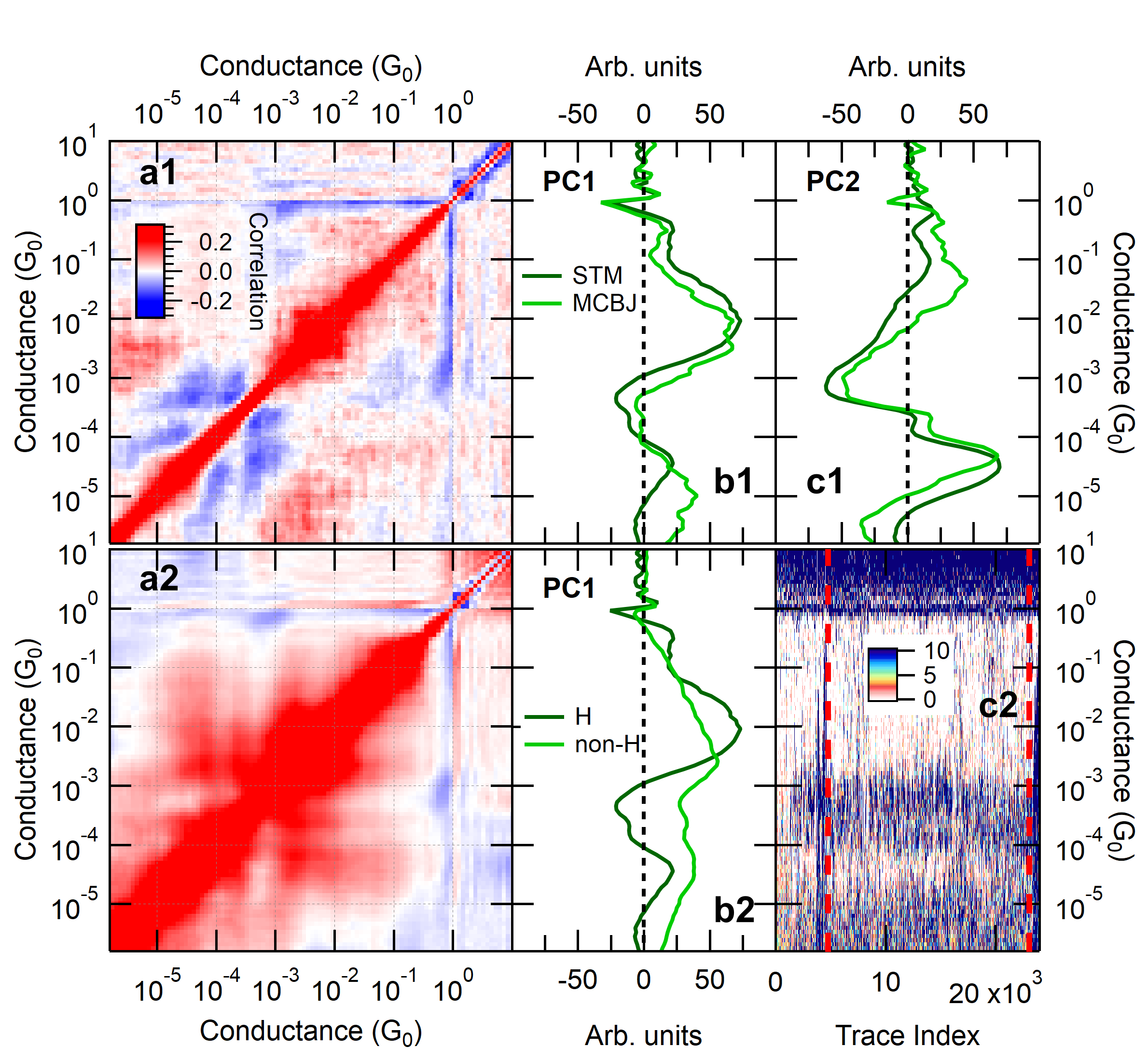}
\caption{The first row compares the correlation plot of an Au-BPY-Au MCBJ dataset (a1) and the corresponding first two principal components (b1,c1) to STM break junction measurements, demonstrating excellent sample-to-sample and method-to-method reproducibility. The second row illustrates that the PCP analysis is extremely sensitive to the homogeneity of the investigated traces. Panel (c2) shows the temporal histogram of the entire Au-BPY-Au STM dataset. If the inhomogeneous parts outside the interval bordered by red dashed lines are included in the analysis, both the correlation plot (a2) and the principal components strongly deviate from the well-established behavior of the Au-BPY-Ay system. The latter is illustrated in panel (b2) comparing $P^{(1)}$ for homogeneous (dark green) and inhomogeneous (light green) datasets. Similar problems arise, if the bottom conductance cutoff is not set properly, and therefore the base noise level of the current amplifier is also included in the conductance region of the analysis (not shown).}
  \label{SFig1}}
\end{figure}

To test the robustness of the PCP analysis, we compare the correlation matrices and the principal components for Au-BPY-Au junctions measured in two independent experiments, using two completely different measurement methods (MCBJ/STM break junctions), see Figure \ref{SFig1}. The correlation matrix of the MCBJ dataset (Fig.~\ref{SFig1}a1) indeed resembles that of the STM dataset (Fig.\;1c in the manuscript) including fine details of the distinct positive/negative correlation lobes. Accordingly, the principal components are also very similar, as illustrated for the first two principal components in Fig.~\ref{SFig1}b1,c1 (see the light/dark green lines for the MCBJ/STM data). This sample-to-sample and method-to-method reproducibility (also confirmed by further measurements on other independent datasets) underpins that the correlation plot and its principal components indeed act as specific fingerprints of the possible single-molecule junction evolution trajectories. 

Figure \ref{SFig1}a2-c2, on the other hand, demonstrate an aspect, where the PCP method is extremely sensitive to the data preparation. Panel c2 shows the so-called temporal conductance histogram,\cite{Magyarkuti2016} where the horizontal axis is the trace index, $r$, and the vertical cuts represent the single-trace conductance histograms, $N_i(r)$ as a colorscale. This plot visualizes the temporal evolution of the conductance traces. For a reliable measurement, the temporal homogeneity of the temporal conductance histogram is required. This is satisfied in the region between the red dashed lines, for which region the correlation matrix was calculated in Fig.\;1c in the manuscript. Before the first dashed line, however, a region is observed, where the molecular structures are missing in the temporal histogram. Such an evident inhomogeneity in the data yields a completely different correlation plot (Figure \ref{SFig1}a2) than the correlation plot for the homogeneous portion (Fig.\;1c in the manuscript), which also affects the principal components. Note, that Figure \ref{SFig1}b2 only compares $P^{(1)}$ for the homogeneous and the non-homogeneous datasets, but all the leading principal components strongly deviate from the well-defined PCs of the homogeneous dataset. This comparison highlights a weak point of the PCP-NN analysis: special care must be taken to ensure the homogeneity of the conductance traces (and to check the results of the analysis for various independent homogeneous datasets), otherwise the results will reflect artifacts due to the data inhomogeneity instead of the well-defined internal correlations.

\section{The basics of the noise measurements}
\begin{figure}[t!]
\centering{\includegraphics[width=0.64\textwidth]{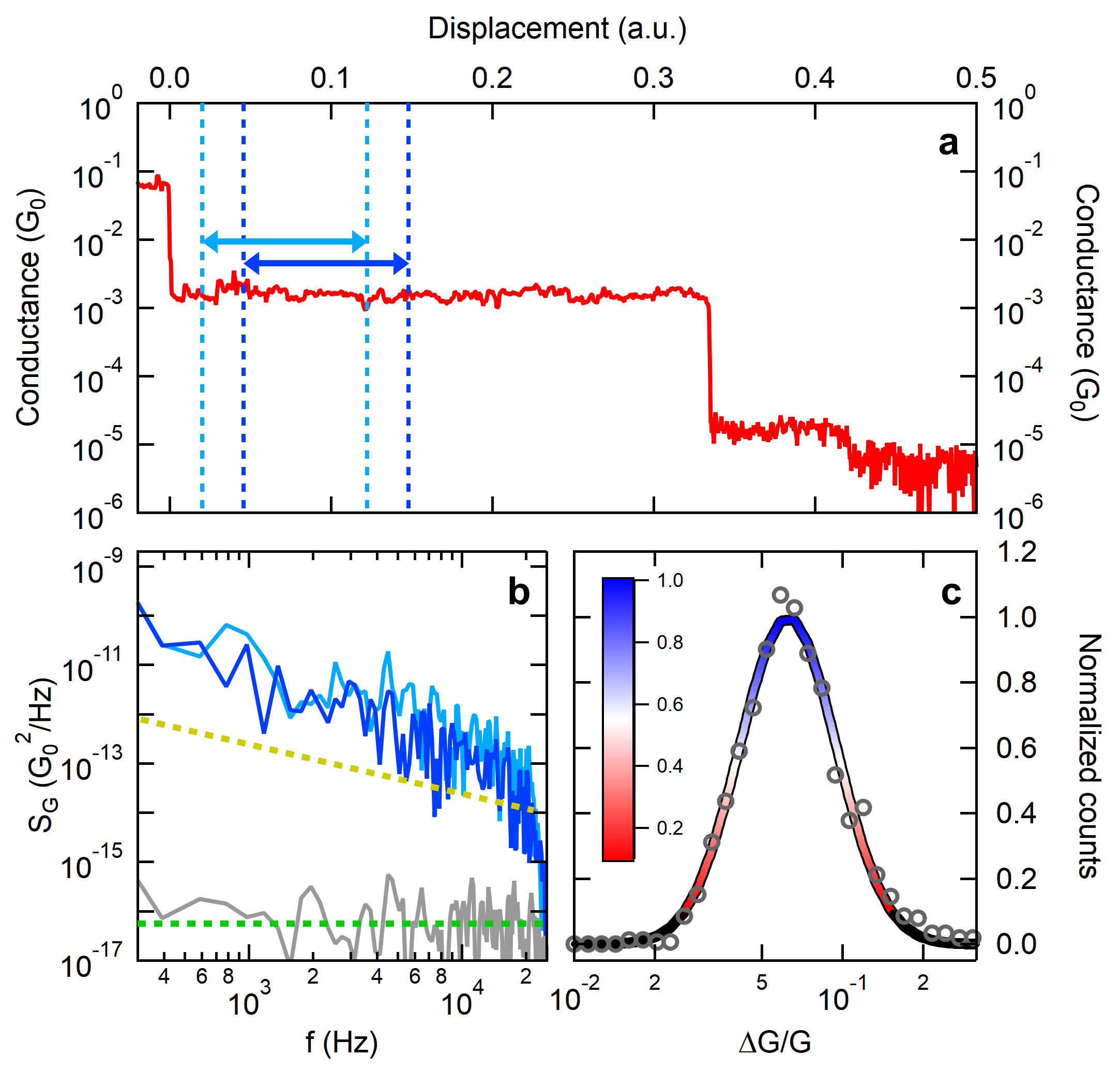}
  \caption{The noise spectra measured at $100\,$mV bias (light and dark blue lines in panel (b)) are calculated by FFT from fixed length moving regions of the conductance traces (light and dark blue regions in (a)). In panel (b) the gray curve shows the significantly smaller noise floor of our measurement system in comparison to the expected thermal noise floor (green dashed line). The yellow dashed line illustrates the $S_G\sim 1/f$ frequency dependence. From the spectra, the relative conductance noise is calculated (see text), and the such obtained $\Delta G/G$ values are gathered to a histogram ($20\,$bins/decade) for all the trace portions exhibiting a certain conductance (panel (c)). Gaussian fitting along the logarithmic x-axis is applied to determine the most probable $\Delta G/G$ value (i.e. the peak position of the Gaussian).}
  \label{SFig2}}
\end{figure}

The $1/f$-type noise analysis\cite{Balogh2021} was carried out on conductance-displacement traces measured via the MCBJ method, using a special measurement system optimized for noise measurements. This includes a 24-bit resolution acquisition board with an anti-aliasing filter (NI-9239) and a low noise current amplifier (FEMTO DLPCA-200). 
Fig.~\ref{SFig2}a illustrates a single-molecule plateau with dominant HighG character on a conductance trace using a sampling rate of $50\,$kHz.  Such conductance plateaus typically contain several thousands of data points. The noise spectra (light and dark blue lines in panel (b)) are calculated by fast Fourier transformation (FFT) for various, 256 datapoints long segments of the plateaus (see the regions denoted by light and dark blue arrows as examples in panel (a)), excluding the unstable trace-segments (i.e. where the conductance ratio between the beginning and the end of the evaluated trace segment is larger than a factor of two). 
Afterward, the noise spectrum is integrated for the $f=2-5\,$kHz frequency band yielding the mean squared conductance variation, $\left<(\Delta G)^2\right>$ in this band. From this the $\Delta G/G=\sqrt{\left<(\Delta G)^2\right>}/\left<G\right>$ relative conductance noise is calculated at various mean conductances $\left<G\right>$. The histogram of the such-obtained $\Delta G/G$ values is plotted for all the trace-segments, for which the $\left<G\right>$ average conductance falls into the same conductance bin (see panel (c)). By Gaussian fitting along the logarithmic horizontal axis the most probable $\Delta G/G$ value can be determined, which can be plotted as the function of the mean conductance. This is demonstrated by the black line in Fig.\;4a in the manuscript representing the most probable relative conductance noise for the entire Au-BPY-Au dataset, as well as by the blue and red lines Fig.\;4b in the manuscript representing the configuration specific noise data. Note, that the color-scale of the Gaussian fit in Fig.~\ref{SFig2}c encodes the ratio of the actual value on the Gaussian compared to the peak value. This color-coding is reproduced in Fig.\;4a in the manuscript demonstrating the scattering of the noise data compared to the most probable noise value.   

\bibliography{PCP_refs}